\documentclass[letterpaper,10pt]{article} 

\usepackage{opticameet3} %

\newcommand\authormark[1]{\textsuperscript{#1}}

\usepackage{amsmath,amssymb}
\usepackage[acronym,nohypertypes=acronym]{glossaries}
\usepackage[colorlinks=true,bookmarks=false,citecolor=blue,urlcolor=blue]{hyperref}
\usepackage{tikz}
\usepackage{pgfplots}
\pgfplotsset{width=7cm,compat=1.3}
\usetikzlibrary{shapes,arrows,positioning,calc, matrix,spy,patterns}
\usetikzlibrary{decorations.pathreplacing,calligraphy}
\usetikzlibrary{colorbrewer}
\usetikzlibrary{fit}
\usepgfplotslibrary{colorbrewer}
\usepackage[nolist]{acronym}
\usepackage{siunitx}
\usepackage{setspace}
\usepackage[font={footnotesize}]{caption} 
\usepackage{stackengine}
\definecolor{kit-green100}{rgb}{0,.59,.51}
\definecolor{kit-green70}{rgb}{.3,.71,.65}
\definecolor{kit-green50}{rgb}{.50,.79,.75}
\definecolor{kit-green30}{rgb}{.69,.87,.85}
\definecolor{kit-green15}{rgb}{.85,.93,.93}
\definecolor{KITgreen}{rgb}{0,.59,.51}

\definecolor{KITpalegreen}{RGB}{130,190,60}
\colorlet{kit-maigreen100}{KITpalegreen}
\colorlet{kit-maigreen70}{KITpalegreen!70}
\colorlet{kit-maigreen50}{KITpalegreen!50}
\colorlet{kit-maigreen30}{KITpalegreen!30}
\colorlet{kit-maigreen15}{KITpalegreen!15}

\definecolor{KITblue}{rgb}{.27,.39,.66}
\definecolor{kit-blue100}{rgb}{.27,.39,.67}
\definecolor{kit-blue70}{rgb}{.49,.57,.76}
\definecolor{kit-blue50}{rgb}{.64,.69,.83}
\definecolor{kit-blue30}{rgb}{.78,.82,.9}
\definecolor{kit-blue15}{rgb}{.89,.91,.95}

\definecolor{KITyellow}{rgb}{.98,.89,0}
\definecolor{kit-yellow100}{cmyk}{0,.05,1,0}
\definecolor{kit-yellow70}{cmyk}{0,.035,.7,0}
\definecolor{kit-yellow50}{cmyk}{0,.025,.5,0}
\definecolor{kit-yellow30}{cmyk}{0,.015,.3,0}
\definecolor{kit-yellow15}{cmyk}{0,.0075,.15,0}

\definecolor{KITorange}{rgb}{.87,.60,.10}
\definecolor{kit-orange100}{cmyk}{0,.45,1,0}
\definecolor{kit-orange70}{cmyk}{0,.315,.7,0}
\definecolor{kit-orange50}{cmyk}{0,.225,.5,0}
\definecolor{kit-orange30}{cmyk}{0,.135,.3,0}
\definecolor{kit-orange15}{cmyk}{0,.0675,.15,0}

\definecolor{KITred}{rgb}{.63,.13,.13}
\definecolor{kit-red100}{cmyk}{.25,1,1,0}
\definecolor{kit-red70}{cmyk}{.175,.7,.7,0}
\definecolor{kit-red50}{cmyk}{.125,.5,.5,0}
\definecolor{kit-red30}{cmyk}{.075,.3,.3,0}
\definecolor{kit-red15}{cmyk}{.0375,.15,.15,0}

\definecolor{KITpurple}{RGB}{160,0,120}
\colorlet{kit-purple100}{KITpurple}
\colorlet{kit-purple70}{KITpurple!70}
\colorlet{kit-purple50}{KITpurple!50}
\colorlet{kit-purple30}{KITpurple!30}
\colorlet{kit-purple15}{KITpurple!15}

\definecolor{KITcyanblue}{RGB}{80,170,230}
\colorlet{kit-cyanblue100}{KITcyanblue}
\colorlet{kit-cyanblue70}{KITcyanblue!70}
\colorlet{kit-cyanblue50}{KITcyanblue!50}
\colorlet{kit-cyanblue30}{KITcyanblue!30}
\colorlet{kit-cyanblue15}{KITcyanblue!15}

\definecolor{KITbraun}{RGB}{167,130,46}

\DeclareMathAlphabet{\mathcal}{OMS}{cmsy}{m}{n}
\usetikzlibrary{external}
\tikzset{lsblock/.style = {rectangle, thick, draw, minimum width=1.4cm, minimum height=0.8cm, rounded corners=1.6mm, font=\footnotesize},}

\newcommand\blfootnote[1]{%
  \begingroup
  \renewcommand\thefootnote{}\footnote{#1}%
  \addtocounter{footnote}{-1}%
  \endgroup
}

\begin{document}

\begin{acronym}[TROLL]
	\acro{BPS}[BPS]{blind phase search}
	\acro{CDC}[CDC]{chromatic dispersion compensation}
	\acro{CPSD}[CPSD]{cross power spectral density}
	\acro{EEPN}[EEPN]{equalization-enhanced phase noise}
	\acro{FIR}[FIR]{finite impulse response}
	\acro{LO}[LO]{local oscillator}
    \acro{MC}[MC]{multi-carrier}
    \acro{MF}{matched filter}
	\acro{SC}[SC]{single-carrier}
	\acro{SNR}[SNR]{signal-to-noise ratio}
\end{acronym}

\title{A Novel Phenomenological Model of Equalization-enhanced Phase Noise}
\author{Benedikt Geiger\authormark{1,*}, Fred Buchali\authormark{2}, Vahid Aref\authormark{2}, and Laurent Schmalen\authormark{1}}
\address{\authormark{1} Communications Engineering Lab (CEL), Karlsruhe Institute of Technology (KIT), 76187 Karlsruhe, Germany\\
\authormark{2}Nokia, Magirusstr. 8, 70469 Stuttgart, Germany}
\email{\authormark{*}\texttt{benedikt.geiger@kit.edu}} %

\vspace{-3mm}
\begin{abstract}
We show that equalization-enhanced phase noise manifests as a time-varying, frequency-dependent phase error, which can be modeled and reversed by a time-varying all-pass finite impulse response filter.
\end{abstract}

\blfootnote{B. Geiger conducted this research during his Master Thesis at Nokia, Stuttgart. This project received funding from the European Research Council (ERC) under the European Union’s Horizon 2020 research and innovation program RENEW (grant agreement No. 101001899).}

\vspace*{-1mm}
\section{Introduction}
\vspace*{-2mm}
Increasing the throughput and reducing the cost per bit of coherent optical communication systems requires increasing the symbol rate per channel. However, a major challenge in such high-baud rate systems is \ac{EEPN}, which arises from the interplay between the phase noise of the \ac{LO} and the digital \ac{CDC} filter~\cite{shieh_equalization-enhanced_2008}. This effect is particularly critical for low-cost systems that use lasers with large linewidth, such as 800G ZR+ transceivers~\cite{jung_mitigating_2024}.

\ac{EEPN} has been mostly studied from a statistical perspective~\cite{martins_frequency-band_2024,xu_study_2024} and is considered as a combination of timing jitter and inter-symbol interference~\cite{jung_mitigating_2024,xu_study_2024,qiu_mitigation_2024}, since it can be partially mitigated using timing error correction~\cite{martins_frequency-band_2024} and adaptive filters~\cite{jung_mitigating_2024,Abol_JLT}. However, a detailed investigation of the distortions caused by \ac{EEPN} is still missing.

In this paper, we present a novel phenomenological and system-theoretical model to describe \ac{EEPN} and the induced distortions. In particular, we use a \ac{MC} transmission system to investigate the frequency-dependent nature of the phase noise and provide a detailed analysis of the distortion caused by \ac{EEPN} in a \ac{SC} transmission. We show that an all-pass filter suffices to describe the distortion and reverse the phase error using an all-pass \ac{FIR} filter.

\vspace*{-1mm}
\section{System Model}
\begin{figure}[!h]
	\begin{center}
		\input{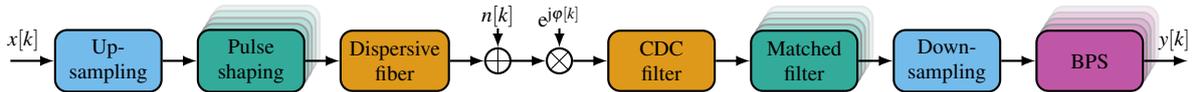}
	\end{center}
    \vspace{-0.6cm}
	\caption{Block diagram of the equivalent baseband model used to investigate \acs{EEPN} in an \ac{SC} and \ac{MC} transmission.}
    \vspace{-0.65cm}
	\label{fig:EEPN_Chain}
\end{figure}

\vspace*{-2mm}
Fig.~\ref{fig:EEPN_Chain} illustrates the equivalent baseband model used to investigate \ac{EEPN} in both \ac{SC} and \ac{MC} systems. First, modulation symbols $x[k]$ are generated at a symbol rate of \mbox{$1/T_{\mathrm{S}}$ = \SI{180}{GBd}}, upsampled by a factor of \mbox{$m \! = \! \num{2}$} and a root-raised cosine filter with a roll-off factor of $\num{0.05}$ is applied. In the case of an \ac{MC} transmission, the transmit symbols are divided into \num{8} parallel streams, where the pulse shaping filter is applied for each stream independently before the streams are multiplexed digitally in the frequency domain. We model a terrestrial, nonlinearity-free transmission over $\SI{6600}{km}$ of Corning TXF fiber at a wavelength of $\SI{1550}{\nano \meter}$. The fiber has a chromatic dispersion of \mbox{$23\,\text{ps}\,\text{nm}^{-1}\,\text{km}^{-1}$} and additive white Gaussian noise $n[k]$ with a \ac{SNR} of $\SI{13}{dB}$ is added to account for accumulated amplified spontaneous emission noise from optical amplifiers. At the coherent receiver, phase noise is added due to the finite linewidth \mbox{$\Delta \nu \! = \! \SI{70}{kHz}$} of the \ac{LO}. The phase noise
\vspace{-0.2cm}
\begin{equation}
    \varphi[k+1] = \varphi[k] + \Delta \varphi[k], \qquad
	\Delta \varphi[k] \sim  \mathcal{N} (0, \sigma^2_{\varphi}), \qquad
	\sigma_{\varphi}^2 = 2\pi \Delta \nu \cdot T_\mathrm{S}/m
    \vspace{-0.2cm}
    \label{eq:unused}
\end{equation}
is modeled as a Wiener process. In the digital domain, a \ac{CDC} and \ac{MF} are applied, the signal is downsampled, and the phase is estimated and corrected using \ac{BPS}. The \ac{MF} and \ac{BPS} are applied to each sub-carrier independently in an \ac{MC} system. We neglect the transmitter phase noise because it passes through the fiber channel and the \ac{CDC} and does not result in \ac{EEPN} if the \ac{CDC} is applied at the receiver~\cite{shieh_equalization-enhanced_2008}.
\vspace{-0.55cm}

\vspace*{-1mm}
\section{Equalization Enhanced Phase Noise is a Frequency-dependent Phase Error}
\begin{figure}[!t]
    \input{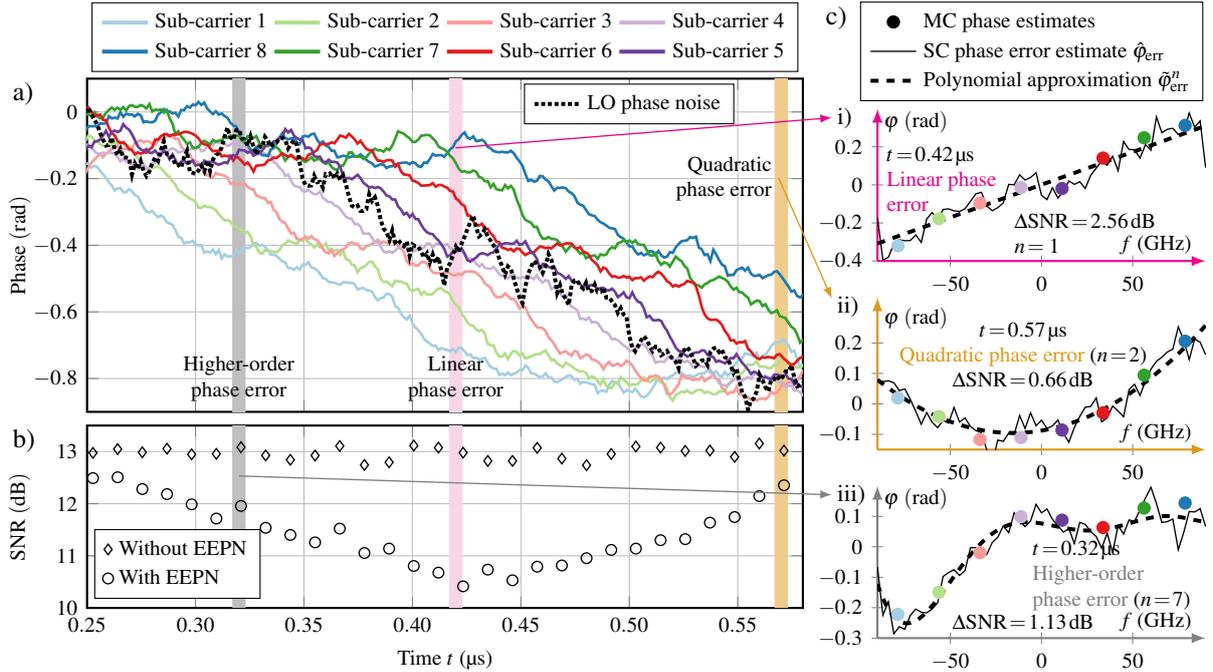}
    \vspace{-0.7cm}
    \caption{a) Estimated phases in an \ac{MC} system, illustrating that \ac{EEPN} manifests as time-varying frequency-dependent phase noise. b) Blockwise \acs{SNR} for an \ac{SC} transmission using the same \ac{LO} phase noise realization as shown in a). c) Frequency-dependent phase (error) in an \ac{SC} and \ac{MC} transmission at i) $t \! = \! \SI{0.42}{\micro \second}$, ii) $t \! = \! \SI{0.57}{\micro \second}$, and iii) $t \! = \! \SI{0.32}{\micro \second}$.}
    \label{fig:phase_noise_MC}
    \vspace{-0.55cm}
\end{figure}

\vspace*{-2mm}
In this section, we present a novel methodology for analyzing \ac{EEPN} and the resulting distortions in an \ac{SC} transmission. Specifically, we utilize an \ac{MC} setup to demonstrate that \ac{EEPN} is a frequency-dependent phase error. Fig.~\ref{fig:phase_noise_MC}a shows a realization of the \ac{LO} phase noise and the estimated phases for each sub-carrier in the \ac{MC} setup. %
Since the \ac{CDC} filter introduces a frequency-dependent time shift to the received signal and \ac{LO} phase noise, the estimated phase noise of each sub-carrier becomes a time-shifted version of the \ac{LO} phase noise, with the time shift depending on the sub-carrier~\cite{qiu_mitigation_2024}. %
More importantly, we observe in Fig.~\ref{fig:phase_noise_MC} that the estimated phase noise depends on the sub-carrier at a fixed point in time, which means that the phase noise is frequency-dependent.

In \ac{MC} systems, the independent phase estimation for each sub-carrier addresses the frequency-dependency of the phase noise, effectively mitigating \ac{EEPN}.
In contrast, an \ac{SC} system employs only a single \ac{BPS}, which is incapable of accounting for the frequency-dependent nature of the phase noise. Since the \ac{LO} phase fluctuates randomly over time, \ac{EEPN} manifests as a time-varying, frequency-dependent phase error. Since the \ac{CDC} filter has a constant magnitude in the frequency domain, we propose to model the \ac{EEPN} distortions that occur in an \ac{SC} transmission as a time-varying all-pass filter where the phase error corresponds to the phase of this filter.

To address the time-varying behavior, we subdivide the signal into blocks of $\SI{2048}{symbols}$ ($\SI{11.4}{ns}$), in which the variation of the \ac{LO} phase noise is small. Fig.~\ref{fig:phase_noise_MC}b) shows the blockwise \ac{SNR} of an \ac{SC} transmission with the same \ac{LO} realization as the \ac{MC} transmission. We observe that the \ac{SNR} penalty in the \ac{SC} transmission increases when the sub-carrier phase estimates in the \ac{MC} system deviate from the \ac{LO} phase noise, due to the resulting phase error in the \ac{SC} transmission, with larger phase excursions leading to higher penalties. %
We determine the frequency-dependent phase error within each block using the phase of the \ac{CPSD}
\vspace{-0.3cm}
\begin{equation}
    \hat{\varphi}_{\mathrm{err}}(f) = \sphericalangle \{ S_{xy}(f) \}, \quad S_{xy}(f) = \int_{-\infty}^{\infty} r_{xy} (\tau) e^{-\mathrm{j} 2 \pi f \tau} \mathrm{d} \tau, \quad r_{xy} (\tau) = \int_{-\infty}^{\infty} x (t + \tau) y^*(t) \mathrm{d} t,
    \label{eq:phase_error}
    \vspace{-0.3cm}
\end{equation}
between transmitted symbols $x$ and estimated symbols $y$, where the \ac{CPSD} is the Fourier transform of the cross-correlation. Fig.~\ref{fig:phase_noise_MC}c) shows three characteristic phase error examples and demonstrates that the calculated phase error~(\ref{eq:phase_error}) closely matches the phase estimates of the \ac{MC} setup. To analyze the distortions in detail, we approximate the phase error by an $n$\textsuperscript{th} order polynomial
\vspace{-0.22cm}
\begin{equation}
    \hat{\varphi}_{\mathrm{err}}(f) \approx \tilde{\varphi}^n_{\mathrm{err}} (f) = \varphi_0 + \varphi_1 \cdot f + \varphi_2 \cdot f^2 + \varphi_3 \cdot f^3 + \ldots + \varphi_n \cdot f^n,
    \label{eq:EEPN:Höhere_Ordning_Phase}
    \vspace{-0.22cm}
\end{equation}
where the coefficients $\varphi_0, \ldots, \varphi_n$ are chosen to minimize the mean squared error between the estimated~(\ref{eq:phase_error}) and approximated~(\ref{eq:EEPN:Höhere_Ordning_Phase}) phase error. \mbox{Fig.~\ref{fig:phase_noise_MC}c)-i)} illustrates that a first-order polynomial approximates the phase error at the maximum \ac{EEPN} penalty of \mbox{$\Delta\mathrm{SNR} \! = \! \SI{2.56}{dB}$} well. We observe that linear phase errors cause the largest phase excursions and penalties due to their constant slope across the signal bandwidth. This linear phase shift corresponds to a timing offset proportional to the slope $\varphi_1$, with a phase change of $\SI{0.62}{rad}$ across the signal bandwidth, resulting in a timing offset of $\SI{9.85}{\%}$ of a unit interval. However, \ac{EEPN} can also introduce higher-order phase errors, which typically result in smaller phase excursions and penalties than linear phase errors. \mbox{Fig.~\ref{fig:phase_noise_MC}c)-ii)} shows a quadratic phase error, approximated by a second-order polynomial, describing a mainly dispersive behavior proportional to $\varphi_2$ with an \ac{SNR} penalty of \mbox{$\Delta \mathrm{SNR} \! = \! \SI{0.66}{dB}$}. The phase error in \mbox{Fig.~\ref{fig:phase_noise_MC}c)-iii)} requires a seventh-order polynomial for sufficient approximation, indicating a superposition of phase errors of various orders, i.e., timing offset + dispersion + higher-order terms, leading to an \ac{SNR} penalty of \mbox{$\Delta \mathrm{SNR} \! = \!  \SI{1.13}{dB}$}.

\vspace*{-1mm}
\section{All-pass FIR Filter for Phase Reversal and Simulation Results}
\vspace*{-2mm}
To validate our novel system-theoretical description of \ac{EEPN} as an all-pass filter, we demonstrate that an all-pass \ac{FIR} filter, which reverses only the frequency-dependent phase error, is sufficient to reverse the distortions caused by \ac{EEPN}. Previous works~\cite{jung_mitigating_2024,Abol_JLT} lacked a precise model of \ac{EEPN} and imposed no restrictions on the amplitude response of the filter leading to more error-prone coefficient updates. First, we calculate the frequency-dependent phase error blockwise~\cite{jung_mitigating_2024,Abol_JLT,martins_frequency-band_2024} using a block size of \SI{2048}{symbols}, which requires knowledge of the transmit symbols. Alternatively, pilots~\cite{Abol_JLT} or the decision of the received symbol~\cite{jung_mitigating_2024} can serve as a reference. Next, we use an all-pass \ac{FIR} filter \mbox{$H_{\mathrm{rev}}(f) = \mathrm{exp} ( -\mathrm{j} \cdot \varphi_{\mathrm{rev}}(f) ) $} with $\num{61}$ taps in the time domain to reverse the phase error, which is sufficient for our setup. We compare two methods: \mbox{$i$) Optimized} timing which reverses only the linear approximation of the phase error \mbox{$\varphi_{\mathrm{rev}} \! \mathrel{\stackon[1pt]{$=$}{$\scriptstyle!$}} \!  \tilde{\varphi}^1_{\mathrm{err}}$}. \mbox{$ii$) Higher-order} phase reversal which reverses the estimated phase error \mbox{$\varphi_{\mathrm{rev}} \! \mathrel{\stackon[1pt]{$=$}{$\scriptstyle!$}} \!  \hat{\varphi}_{\mathrm{err}}$}. We note that the filter coefficients must be updated for every block in both methods.

\begin{figure}[!t]
            \input{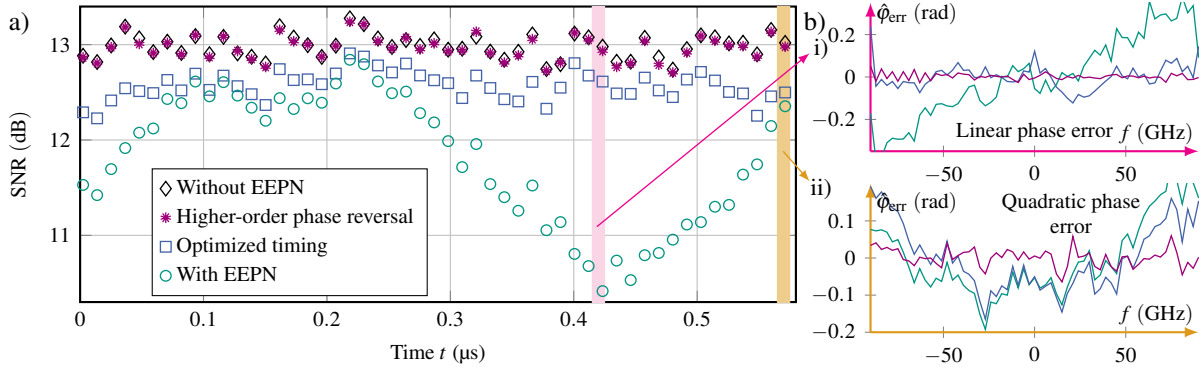}
        \vspace{-0.7cm}
    \caption{Blockwise \ac{SNR} (a) and frequency-dependent phase error (b) after optimized timing and higher-order phase reversal for a i) linear phase error at $t \! = \! \SI{0.42}{\micro \second}$ and a ii) quadratic phase error at $t \! = \! \SI{0.57}{\micro \second}$. The colors used in b) correspond to the ones in a).}
    \label{fig:FIR}
    \vspace{-0.7cm}
\end{figure}

Fig.~\ref{fig:FIR}a) and b) show the blockwise \ac{SNR} and frequency-dependent phase error after optimized timing and higher-order phase reversal. In Fig.~\ref{fig:FIR}b)-i), \ac{EEPN} manifests as a linear phase error, i.e., timing offset, and optimized timing can effectively reverse the phase error, reducing the \ac{SNR} penalty from $\SI{2.56}{dB}$ to $\SI{0.36}{dB}$. Since the large penalties are primarily caused by linear phase errors, optimized timing can reduce the strong drops in \ac{SNR}, where oversampling is necessary for large timing offsets~\cite{qiu_mitigation_2024}. However, \ac{EEPN} introduces also higher-order phase errors, which timing optimization cannot fully reverse because it can only change the slope of the phase error. This limitation is evident in the residual phase error and the remaining \ac{EEPN} penalty observed in Fig.~\ref{fig:FIR}b)-i) and is even more pronounced for the quadratic phase error, i.e. dispersion, in Fig.~\ref{fig:FIR}b)-ii). In both cases, an all-pass filter with more degrees of freedom is necessary to reverse the phase error. We observe that the higher-order phase reversal reduces the maximum phase error to less than $\SI{0.06}{rad}$ in both cases. Thus, the higher-order phase reversal all-pass \ac{FIR} filter decreases the maximum \ac{SNR} penalty from $\SI{2.56}{dB}$ to $\SI{0.08}{dB}$, thereby validating our proposed system-theoretical description of \ac{EEPN} as an all-pass filter.

\vspace*{-1mm}
\section{Conclusion}
\vspace*{-2mm}
We have proposed a novel phenomenological and system-theoretical model of \ac{EEPN}. We showed that an \ac{MC} setup is a suitable means to study \ac{EEPN}, revealing that \ac{EEPN} introduces a time-varying frequency-dependent phase error in an \ac{SC} transmission. This phase error can manifest as a phase offset, timing offset, or higher-order phase distortion, and can be accurately modeled as an all-pass filter. Consequently, a time-varying all-pass \ac{FIR} filter suffices to reverse the \ac{EEPN} distortions and reduce the maximum \ac{EEPN} penalty from \SI{2.56}{dB} to \SI{0.08}{dB}.

\end{document}